\documentclass[twocolumn,english,aps,reprint,twocolumn,superscriptaddress]{revtex4}
\usepackage[T1]{fontenc}
\usepackage[latin9]{inputenc}
\usepackage{color}
\usepackage{amsbsy}
\usepackage{amstext}
\usepackage{amssymb}
\usepackage{graphicx}
\usepackage{esint}

\makeatletter
\@ifundefined{textcolor}{}
{%
 \definecolor{BLACK}{gray}{0}
 \definecolor{WHITE}{gray}{1}
 \definecolor{RED}{rgb}{1,0,0}
 \definecolor{GREEN}{rgb}{0,1,0}
 \definecolor{BLUE}{rgb}{0,0,1}
 \definecolor{CYAN}{cmyk}{1,0,0,0}
 \definecolor{MAGENTA}{cmyk}{0,1,0,0}
 \definecolor{YELLOW}{cmyk}{0,0,1,0}
}

\usepackage[english]{babel}
\def\urlprefix{}
\def\url#1{}
\addto\captionsenglish{%
}

\usepackage{babel}

\usepackage{babel}

\usepackage{babel}

\makeatother

\usepackage{babel}
\begin{document}

\title{Planar Hall effect in antiferromagnetic MnTe thin films}

\author{Gen Yin\footnotemark[1]\footnotemark[2]}

\affiliation{Department of Electrical and Computer Engineering, University of
California -- Los Angeles, Los Angeles, California 90095, USA.}

\author{Jie-Xiang Yu}

\thanks{These two authors contributed equally.}

\affiliation{Department of Physics and Materials Science Program, University of
New Hampshire, Durham, New Hampshire 03824, USA}

\author{Yizhou Liu}

\affiliation{Institute of Physics, Chinese Academy of Sciences, Beijing, China
100190}

\affiliation{Laboratory for Terascale and Terahertz Electronics (LATTE), Department
of Electrical and Computer Engineering, University of California --
Riverside, Riverside, California 92521, USA}

\author{Roger K. Lake}

\affiliation{Laboratory for Terascale and Terahertz Electronics (LATTE), Department
of Electrical and Computer Engineering, University of California --
Riverside, Riverside, California 92521, USA}

\author{Jiadong Zang}

\thanks{Correspond to: genyin@ucla.edu, jiadong.zang@unh.edu, wang@ee.ucla.edu}

\affiliation{Department of Physics and Materials Science Program, University of
New Hampshire, Durham, New Hampshire 03824, USA}

\author{Kang L. Wang}

\thanks{Correspond to: genyin@ucla.edu, jiadong.zang@unh.edu, wang@ee.ucla.edu}

\affiliation{Department of Electrical and Computer Engineering, University of
California -- Los Angeles, Los Angeles, California 90095, USA.}

\affiliation{Department of Physics and Astronomy, University of California --
Los Angeles, Los Angeles, California 90095, USA. }
\begin{abstract}
We show that the spin-orbit coupling (SOC) in $\alpha$-MnTe impacts
the transport behavior by generating an anisotropic valence-band splitting,
resulting in four spin-polarized pockets near $\Gamma$. A minimal
$k\cdot p$ model is constructed to capture this splitting by group
theory analysis, a tight-binding model and \emph{ab initio} calculations.
The model is shown to describe the rotation symmetry of the zero-field
planer Hall effect (PHE). The PHE originates from the band anisotropy
given by SOC, and is quantitatively estimated to be $25\%\sim31\%$
for an ideal thin film with a single antiferromagnetic domain. 
\end{abstract}
\maketitle
Antiferromagnets (AFMs) have been considered as a promising candidate
for next-generation spintronic devices due to their scalability, their
robustness against external magnetic fields, and their ultra-fast
spin dynamics\citep{baltz_antiferromagnetic_2018,gomonay_o._concepts_2017,gomonay_spintronics_2014,jungwirth_antiferromagnetic_2016,zelezny_spin-transport_2017,jungfleisch_perspectives_2018,zelezny_spin_2018}.
Without a net magnetization, conventional means to detect and manipulate
a ferromagnetic order usually cannot be directly employed in AFMs,
which hinders their device applications. Recently, spin-orbit coupling
(SOC) was experimentally shown to enable the detection and manipulation
of the N\'eel-vector orientation in easy-plane AFMs\citep{wadley_electrical_2016,kriegner_multiple-stable_2016,bodnar_writing_2018,grzybowski_imaging_2017},
and therefore became the centerpiece of antiferromagnetic spintronics.
SOC is known to induce spin mixing and band splitting, leading to
unique magnetotransport signatures. The locking between
electron spin and momentum under SOC results in uniform spin accumulation:
the spin-galvanic effect \citep{ganichev_spin-galvanic_2002}. This
effect in some antiferromagents has been shown to exert opposite spin-orbital
torques on anti-parallel local spins, and thereby switches the N\'eel
vector\citep{wadley_electrical_2016,bodnar_writing_2018,grzybowski_imaging_2017}. 

SOC can also lead to anisotropic magnetoresistance (AMR) and planar
Hall effect (PHE)\citep{thomson_xix._1857,smit_magnetoresistance_1951,mcguire_anisotropic_1975,wu_probing_2017,nandy_chiral_2017,burkov_giant_2017}.
In ferromagnetic transition metals and alloys, PHE is known to result
from the $s\textrm{-}d$ mixing given by SOC\citep{smit_magnetoresistance_1951,campbell_spontaneous_1970}.
Although the exact outcome of SOC is strongly material dependent,
AMR and PHE are usually proportional to $\left(\mathbf{M}\cdot\mathbf{j}\right)^{2}$,
and therefore occur in both ferromagnets (FMs) and AFMs\citep{baltz_antiferromagnetic_2018}.
These effects have been experimentally demonstrated in many metallic
and semiconducting AFMs, and are considered as a robust method to
read out the information encoded in the antiferromagnetic order\citep{grzybowski_imaging_2017,wadley_electrical_2016,bodnar_writing_2018,jungwirth_multiple_2018}.

Among many AFMs, $\alpha$-MnTe is particularly attractive both in
terms of fundamental physics and device applications\citep{wasscher_electrical_1969}.
Bulk $\alpha$-MnTe is a p-type semiconductor with
a N\'eel temperature of $T_{N}\approx310\thinspace\textrm{K}$\citep{efrem_dsa_low-temperature_2005,kriegner_magnetic_2017,szuszkiewicz_spin-wave_2006,walther_ultrasonic_1967}.
Due to the semiconducting nature, it is convenient to engineer the
band alignment and the position of the Fermi level. The N\'eel vector
has three coplanar easy axes, which can naturally encode 3-state digital
information\citep{kriegner_magnetic_2017}. In the case of multi-domain,
the most populated N\'eel-vector direction can be easily rotated by
either field cooling or an applied magnetic field as small as $3\thinspace\textrm{T}$\citep{kriegner_multiple-stable_2016}.
These advantages make $\alpha$-MnTe an attractive candidate material
and a convenient building block for antiferromagnetic devices and
other related studies.

In this letter, we seek to theoretically understand the SOC in $\alpha$-MnTe,
and to construct a minimal model that captures the magnetotransport
behavior in the case of a thin film. The zero-field PHE is shown to
originate from the valence-band anisotropy near $\Gamma$ induced
by SOC. The PHE percentage is shown to maximize above the band crossing,
and is estimated to be $25\%\sim31\%$ by a semi-classical transport
calculation based on \emph{ab-initio} bands.

\begin{figure}
\begin{centering}
\includegraphics[width=0.9\columnwidth]{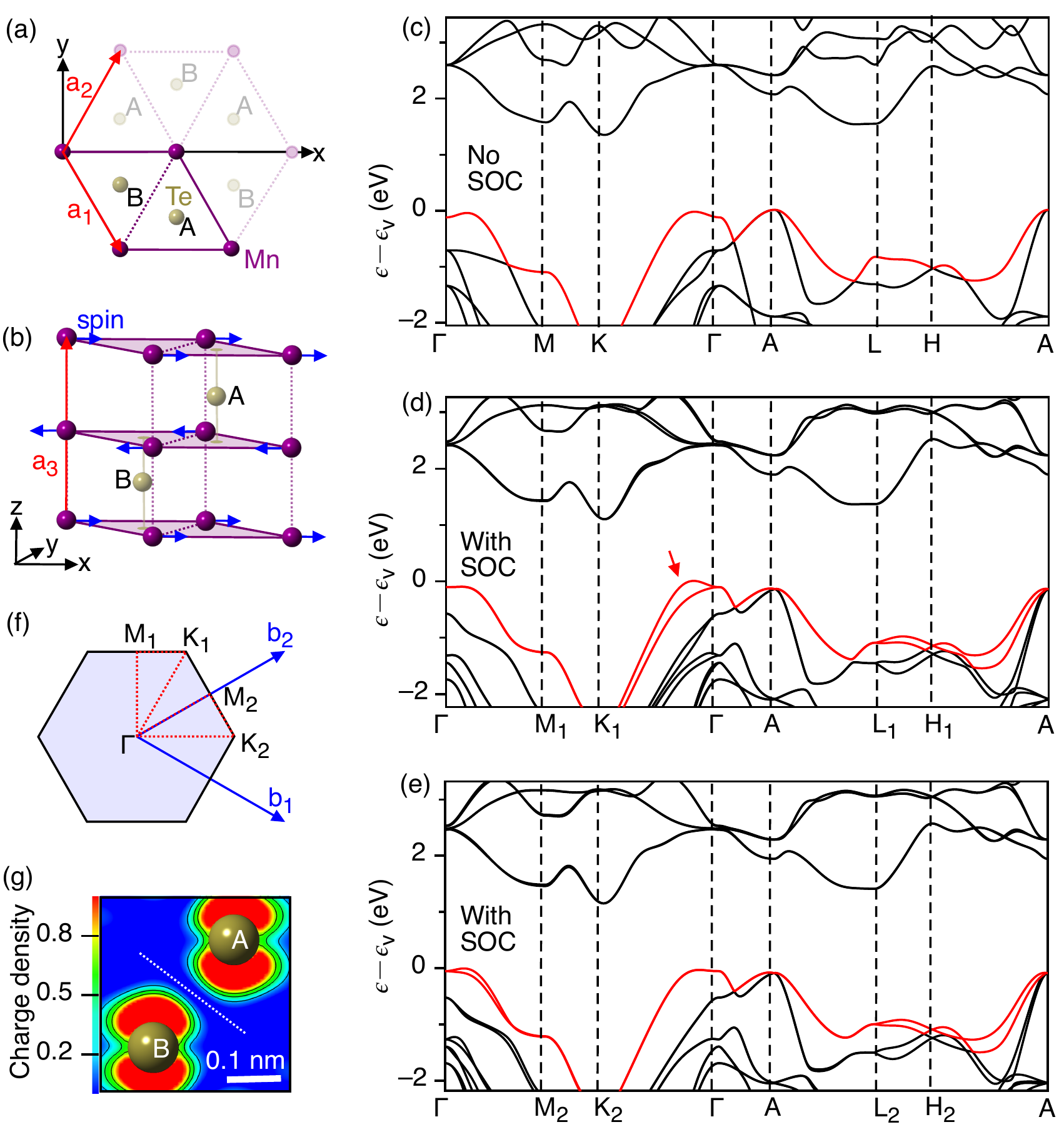} 
\par\end{centering}
\caption{\emph{Ab-initio} band structure for $\alpha$-MnTe. (a) Top view of
the magnetic unit cell and the choice of coordinate. (b) Three dimensional
view of the magnetic unit cell. (c) Electron bands without spin-orbit
coupling (SOC), illustrated along the high-symmetry points of the
non-magnetic primitive cell. The red curve denotes the spin-degenerate
valence band. (d and e) The band structure considering SOC, illustrated
along two different loops in the Brillouin zone shown in (f). The
red arrow denotes the valence band top near $\Gamma.$ (g) The charge
density contributed by the valence band at $\Gamma$. The white dotted
line denotes the plane separating the A-Te and B-Te atoms, where the
charge density is zero. This color contour plot is illustrated in
the plane containing $\hat{a}_{3}$ and the two Mn sublattices. \label{fig:Ab-initio-band}}
\end{figure}

The ground-state magnetic order of $\alpha$-MnTe and the band structure
are captured by first-principles calculations. $\alpha$-MnTe has
a typical NiAs atomic structure as shown in Fig. \ref{fig:Ab-initio-band}(a)
and (b). The lattice constant is relaxed to $a_{1,2}=4.090\thinspace\textrm{\AA}$
and $a_{3}=6.430\thinspace\textrm{\AA}$, $\sim1\%$ smaller than the values observed in
X-ray diffraction \citep{kriegner_multiple-stable_2016}. \textcolor{black}{We will use the relaxed values for the rest of the paper. The impact of the lattice constants will be discussed in Supplementary Materials Sec. I.} Each Mn atom possesses
a local spin moment of $4.40\thinspace\mu_{B}$, indicating $S=\frac{5}{2}$
high spin state, which consists with previous studies\citep{walther_ultrasonic_1967,krause_structural_2013,szuszkiewicz_spin-wave_2006}.
These spins are known to align ferromagnetically within each Mn layer,
whereas the layers stack antiferromagnetically along $\hat{z}$ {[}direction
(001){]}. The antiferromagnetic phase is found to be $774\thinspace\textrm{meV}$
lower in energy than the ferromagnetic phase, indicating an interlayer
antiferromagnetic order as the ground state. 

The band structure near the valence band top is strongly affected
by SOC. As shown in Fig. \ref{fig:Ab-initio-band}(c-e), the valence
band top is found to be at A point without SOC, which is $\sim0.05\thinspace\textrm{eV}$
higher than the $\Gamma$ point. This is consistent with the pioneering
calculations done by Podg\`orny \emph{et al.} \citep{podgorny_electronic_1983}
and Wei \emph{et al.} \citep{wei_total-energy_1987}. Once SOC is
included, the configuration with in-plane spin along $\hat{x}$ {[}shown
in Fig.\ref{fig:Ab-initio-band}(b){]} or other two equivalent directions
has the lowest energy, suggesting that the easy axes are consistent
with the recent experiment\citep{kriegner_multiple-stable_2016}.
With this magnetic order, $C_{3}$ rotation about $\hat{z}$ is no
longer a symmetry operation so that the $\Gamma$-M-K-$\Gamma$ paths
are not identical. This will be explained in-detail by the group-theory
analysis later. Two representative paths are chosen to demonstrate
the band anisotropy, as shown in Fig. \ref{fig:Ab-initio-band}(f).
The most significant SOC-splitting occurs in the valence band near
$\Gamma$-point, as denoted by the red arrow in Fig. \ref{fig:Ab-initio-band}(d).
This splitting shifts the band top from the A point to the $\Gamma\rightarrow$$\textrm{K}_{1}=\left(-\frac{1}{3},\frac{2}{3},0\right)$
line, which is now $\sim0.1\thinspace\textrm{eV}$ higher. Cryogenic
magnetotransport therefore should be dominated by this band, which
is formed by the anti-bonding of the $p_{z}$ orbitals of Te($5p$)
sitting on different sub-lattices, as illustrated by the partial charge
density in Fig. \ref{fig:Ab-initio-band}(g). No band splitting shows
up in the conduction band, which is dominated by the empty Mn($3d$)
$3z^{2}-r^{2}$ orbital. The above calculations are carried out using
project augmented wave pseudo-potential (PAW) \citep{blochl_projector_1994}
implemented in VASP\citep{kresse_efficiency_1996,kresse_efficient_1996}.
Generalized gradient approximation (GGA) in Perdew, Burke, and Ernzerhof
(PBE) \citep{perdew_generalized_1997} is used as the exchange-correlation
energy for structure optimization, whereas hybrid functional (HSE06)
is applied for the calculation of the total energy. This functional
computes the exact Fock energy and is known to avoid underestimation
of band gaps in certain systems\citep{heyd_hybrid_2003,krukau_influence_2006}.
See Sec. II in Supplemental Material for the comparison
between the calculated band structure and experimental data from different
sources. The $k$-points are sampled on a $\text{\ensuremath{\Gamma}}$-centered
$13\times13\times8$ mesh, and an energy-cutoff of $400\thinspace\textrm{eV}$
is used throughout all calculations.

\begin{figure}
\begin{centering}
\includegraphics[width=1\columnwidth]{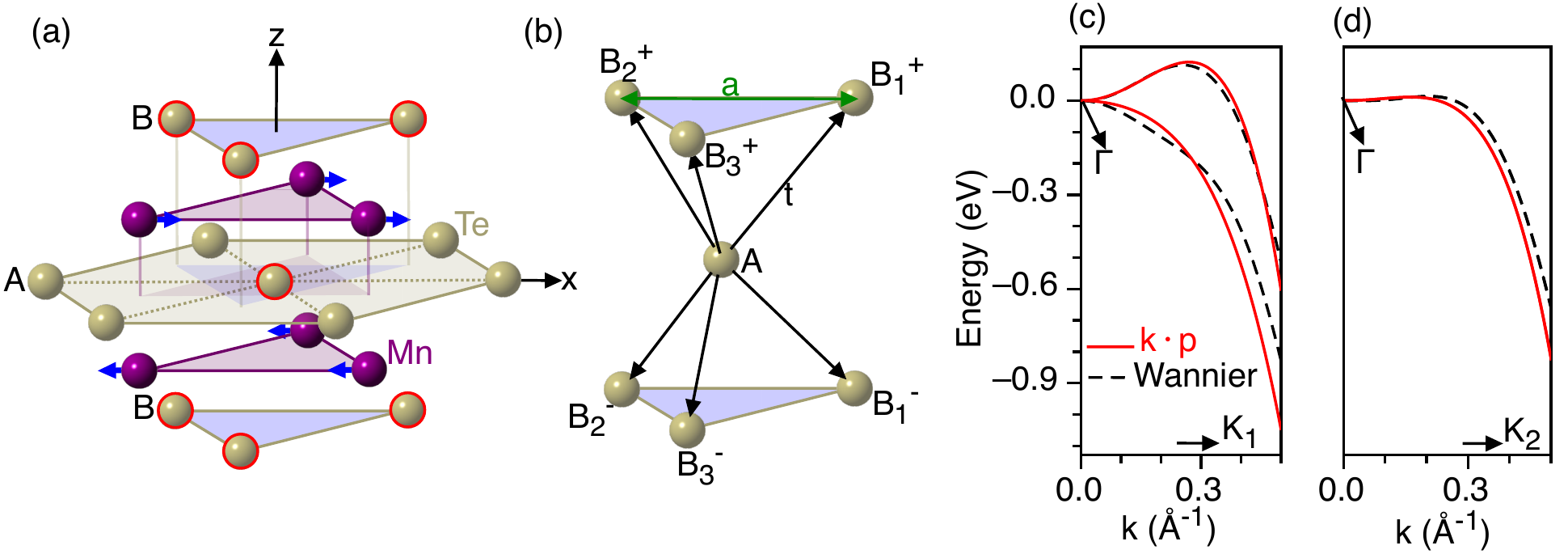} 
\par\end{centering}
\caption{(a) Atoms that are close to the central Te for a bulk $\alpha$-MnTe.
The red circles label the first-nearest Te atoms used in the tight-binding
model. (b) The atomic structure used for the tight-binding model.
Each A-site Te is surrounded by six first-nearest B-site ones. (c)
and (d), the comparison between the $k\cdot p$ model and the Wannier
model from the\emph{ ab initio} calculation. The parameters are $k_{0}=0.162\thinspace\textrm{\AA}^{-1}$,
$\alpha=-6.62$, $\beta_{1}=0.202$, $\beta_{2}=1.191$ and $\frac{k_{0}^{2}}{2m}=0.023\thinspace\textrm{eV}$,
as defined in Eq. \ref{eq:TheKP_Hamiltonian}. The comparison is illustrated
along the paths $\Gamma\rightarrow K_{1}$ and $\Gamma\rightarrow K_{2}$
as denoted in Fig. \ref{fig:Ab-initio-band}(f). \label{fig:KP_comparison_TB_structure}}
\end{figure}

To analytically understand the impact of SOC, a minimal effective
Hamiltonian describing the long-wavelength behavior is constructed.
The NiAs structure of $\alpha$-MnTe has the space group $P_{6_{3}/mmc}$
(No. 194). Therefore, the point-group symmetry should be $D_{6h}$\citep{sandratskii_energy_1981}
without considering the magnetic order. However, once an in-plane
easy axis is selected by the Mn spin, the $C_{3}$ symmetry about
$\hat{z}$ is broken, and the point group $D_{6h}$ is reduced to
its subgroup $D_{2h}$. This group contains inversion ($I$) and three
mirror operations with respect to $xy$, $yz$, and $zx$ planes,
respectively. The combination of inversion and mirror leads to three
$C_{2}$ operations with respect to $x$, $y$, and $z$ axes, respectively.
Double group of $D_{2h}$ has $10$ irreducible representations, grouped
into $5$ pairs with opposite parities. The character table of these
representations is shown in Supplemental Material Sec.
III. Since the valence band is formed by the anti-bonding between
two $p_{z}$ orbitals of Te, basis $\left|\phi_{1}\right\rangle =\frac{1}{\sqrt{2}}\left(p_{zA}+p_{zB}\right)\left|\uparrow\right\rangle $
and $\left|\phi_{2}\right\rangle =\frac{1}{\sqrt{2}}\left(p_{zA}+p_{zB}\right)\left|\downarrow\right\rangle $
expand a $\Gamma_{5}^{+}$ irreducible representation of $D_{2h}$,
where the superscript `$+$' denotes the even parity.

The effective Hamiltonian in this sub-Hilbert space can be constructed
by the theory of invariants \citep{winkler_spin-orbit_2003}. Given
$\Gamma_{5}^{+}\times\Gamma_{5}^{+}=\Gamma_{1}^{+}\oplus\Gamma_{2}^{+}\oplus\Gamma_{3}^{+}\oplus\Gamma_{4}^{+}$,
$\hat{H}\left(\mathbf{k}\right)=\sum_{\gamma}a_{\gamma}\sum_{k=1}^{|\Gamma_{\gamma}|}h_{k}^{\gamma}\left(\mathbf{k}\right)\left(\sum_{i,j=1}^{2}C_{ij,k}^{\gamma}\left|\phi_{i}\right\rangle \left\langle \phi_{j}\right|\right)$,
where $h_{k}^{\gamma}\left(\mathbf{k}\right)$ and $|\Gamma_{\gamma}|$
are the basis and dimension of representation $\Gamma_{\gamma}$,
respectively. Coefficients $\bigl\{ a_{\gamma}\bigr\}$ are free parameters
that cannot be dictated from the symmetry analysis. $C_{ij,k}$ are
the Clebsh-Gordan (CG) coefficients available in \citep{koster_space_2012}.
The lowest order basis of $\Gamma_{2}^{+}$ and $\Gamma_{4}^{+}$
are $k_{z}k_{x}$ and $k_{y}k_{z}$, respectively. Because
we are focusing on the transport signature in MnTe thin film, $z$-direction
is modeled as a quantum well state, in which $\langle k_{z}\rangle=0$
and $\langle k_{z}^{2}\rangle=(n\pi/d)^{2}=\textrm{const.}$, where
$d$ is the film thickness and $n$ are integer values labeling different
quantum well states. Since the basis above and their higher order
representations all contain odd orders of $k_{z}$, $\Gamma_{2}^{+}$
and $\Gamma_{4}^{+}$ do not contribute. Up to the fourth order of
momentum, relevant basis of $\Gamma_{3}^{+}$ are $k_{x}k_{y}$, $k_{x}^{3}k_{y}$,
$k_{x}k_{y}^{3}$ and $k_{x}k_{y}k_{z}^{2}$, where again the last
term can be combined with the first one, treating $k_{z}^{2}$ as
a constant. The CG coefficients are $C_{ij,1}^{3}=(\sigma_{z}){}_{ij}$.
For $\Gamma_{1}^{+}$, the CG coefficients are $C_{ij,1}^{1}=I_{ij}$,
whose corresponding Hamiltonian is thus spin-independent. The magnetic
order is now fully included. One should use $\mathbf{k}^{2}$ and
$\mathbf{k}^{4}$, the basis of the $\Gamma_{1}^{+}$ representation
of $D_{6h}$ instead. Anisotropic spinless dispersion appears since
the 6-th order of $\mathbf{k}$, which is neglected. As a result,
we have the generic effective Hamiltonian given by 
\begin{equation}
\hat{H}(\mathbf{k})=\frac{k_{0}^{2}}{2m}\left[\bar{\mathbf{k}}^{2}-\frac{1}{2}\bar{\mathbf{k}}^{4}-\sigma_{z}\left(\alpha\bar{k}_{x}\bar{k}_{y}+\beta_{1}\bar{k}_{x}^{3}\bar{k}_{y}+\beta_{2}\bar{k}_{y}^{3}\bar{k}_{x}\right)\right],\label{eq:TheKP_Hamiltonian}
\end{equation}
where $k_{0}$, $m$, $\alpha$, $\beta_{1}$ and $\beta_{2}$ are
free parameters. The dimensionless momentum $\bar{\mathbf{k}}$ is
defined as $\bar{\mathbf{k}}=\mathbf{k}/k_{0}$, where $k_{0}$ sets
the length scale. The minus sign of the quartic term is consistent
with the first-principles facts that the splitting occurs in the valence
band. To reveal the microscopic origin of this effective Hamiltonian,
a tight-binding model based on the $5p$ orbitals of Te atoms is established.
Each Te atom on A site is surrounded by 6 B-site nearest neighbors
as shown in Fig. \ref{fig:KP_comparison_TB_structure}(a) and (b).
Twelve localized atomic orbitals are used to describe the degrees
of freedom given by two electron spins, two sub-lattices and three
$p_{x,y,z}$ orbitals. The SOC is included by taking $H^{SO}=\lambda\mathbf{L}\cdot\mathbf{S}$
as a perturbative term. The effective Hamiltonian is obtained by a
canonical transformation, expanding up to the first order of $\lambda$.
Keeping $k$ to the fourth order, the resulting effective Hamiltonian
is consistent with Eq. \ref{eq:TheKP_Hamiltonian}. See
Sec. IV in Supplemental Material for the details of the $k_{z}$ quantization
in thin films. The details of the tight-binding model can be found
in Sec. V. 

\begin{figure}
\begin{centering}
\includegraphics[width=1\columnwidth]{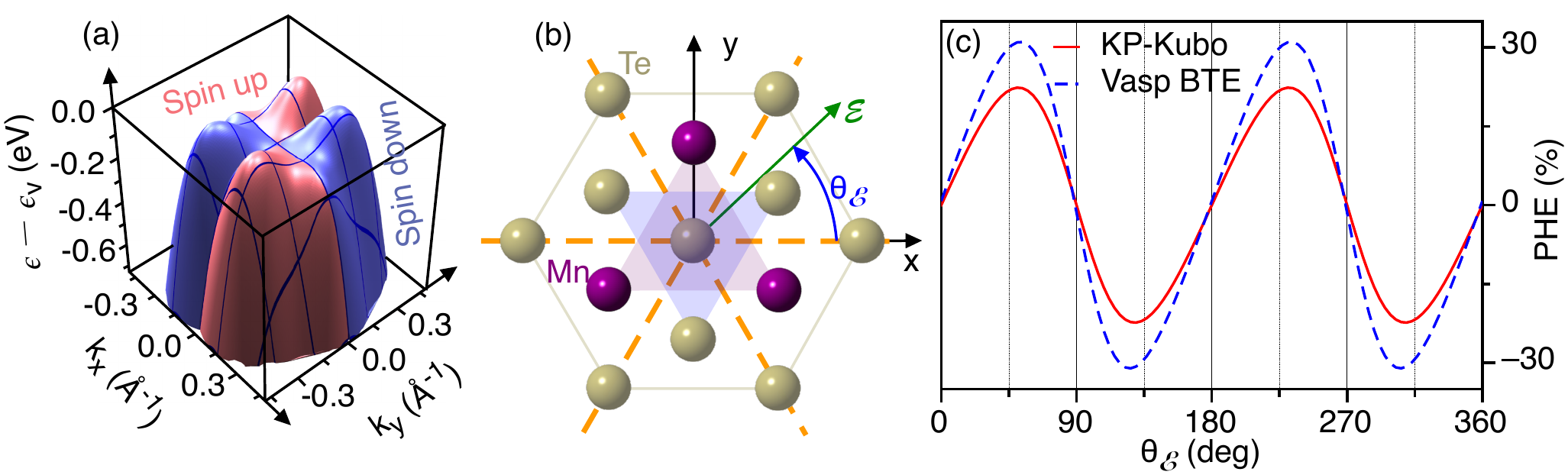} 
\par\end{centering}
\caption{Planar Hall effect percentage as a function of $\theta_{{\cal E}}$,
the angle between the uniform current and the in-plane N\'eel vector.
(a) The full 3D band structure near $\Gamma$ obtained from the Wannier
model. (b) Illustration of the N\'eel vector (along $\hat{x}$), the
applied electric field (green arrow) and three in-plane easy axes
(orange dashed line). This is a top view of Fig. \ref{fig:KP_comparison_TB_structure}(a).
(c) The Hall percentage obtained from transport calculations. \label{fig:Planner-Hall-effect} }
\end{figure}

The parameters in Eq. \ref{eq:TheKP_Hamiltonian} can be obtained
by fitting the $k\cdot p$ model to the \emph{ab-initio} bands. Here,
the fitting target is obtained by transferring the plane-wave basis
obtained by VASP into Wannier function basis, resulting in a Hamiltonian
of localized atomic orbitals (implemented by Wannier90)\citep{mostofi_updated_2014,souza_maximally_2001}.
The fitting parameters are obtained by machine learning using non-linear
conjugate gradient regression and golden-section line search, minimizing
the fitting errors near $\Gamma$. See Sec. VI in
Supplemental Material for the details of the machine-learning algorithm.
With the optimized parameters, the $k\cdot p$ bands are compared
to the \emph{ab-initio} bands along $\Gamma\rightarrow\textrm{K}_{1}$
and $\Gamma\rightarrow\textrm{K}_{2}$, as shown in Fig. \ref{fig:KP_comparison_TB_structure}(c)
and (d), respectively. The spin texture of the valence band is shown
in Fig. \ref{fig:Planner-Hall-effect}(a), where the band edge splits
into four pockets polarized along $\pm z$. Importantly, unlike the
conventional spin-orbit coupling, the spin-dependent term here is
quadratic or quartic in momentum, such that neither time reversal
${\cal T}$ nor its combination with fractional translation (${\cal T}T_{1/2}$)
is a symmetry of the lattice. This therefore leads to nonzero Hall
conductivity even in the absence of external magnetic field.

To capture the zero-field PHE, transport behavior induced by the valence-band
splitting is studied. Scattering centers induced by vacancies of Mn
atoms are considered. Magnetic moments of these impurities point in
$\eta\hat{x}$ directions, where $\eta=\pm1$, denoting two sub-lattices.
The impurity potential is thus written as $\hat{V}=\sum_{\{i,\eta\}}v_{0}\left(a+b\eta\sigma_{x}\right)\delta\left(r-R_{i\eta}\right)$
where $R_{i\eta}$ are positions of magnetic impurities, whereas $a$
and $b$ are spin-independent and -dependent scatterings, respectively.
The current operator along any direction $\hat{n}$, $j=\hat{n}\cdot(\partial\hat{H}/\partial\mathbf{k})$,
is diagonal so that no inter-band transition occurs. Intrinsic Berry
phase contribution is thus absent. In the diffusive regime with low
impurity concentration, the conductivity can be derived by the Kubo-Streda
formula 
\begin{equation}
\sigma_{\perp\left(\parallel\right)}^{\theta_{{\cal E}}}=\frac{\hbar}{2\pi}\textrm{Tr}\left[j_{\parallel}^{\theta_{{\cal E}}}G^{R}\left(\epsilon_{F}\right)j_{\perp\left(\parallel\right)}^{\theta_{{\cal E}}}G^{A}\left(\epsilon_{F}\right)\right]\label{eq:KuboFormula}
\end{equation}
where $j_{\parallel}$ and $j_{\perp}$ are current operators in the
parallel and perpendicular directions with respect to the electric
field direction, as shown in Fig. \ref{fig:Planner-Hall-effect}(b).
Given $\theta_{{\cal E}}$ as the angle between the electric field
$\boldsymbol{{\cal E}}$ and $\hat{x}$, $j_{\parallel}^{\theta_{{\cal E}}}=j_{x}\cos\theta_{{\cal E}}+j_{y}\sin\theta_{{\cal E}}$,
and $j_{\perp}^{\theta_{{\cal E}}}=-j_{x}\sin\theta_{{\cal E}}+j_{y}\cos\theta_{{\cal E}}$.
$G^{R,A}=[(G_{0}^{R,A})^{-1}-\Sigma^{R,A}]^{-1}$ is the retarded
(advanced) Green function. In Born approximation, $\Sigma^{R,A}=\langle\hat{V}\rangle+\langle\hat{V}G^{R,A}\hat{V}\rangle$,
where $\langle\cdots\rangle$ is the impurity average. Assuming the
two Mn sub-lattices have the same impurity concentration $n_{i}$,
one obtains $\langle\hat{V}\rangle=2an_{i}v_{0}$, which is just a
constant absorbed by the Fermi energy. Therefore, $\Sigma^{R,A}=\mp i/2\tau$,
with the relaxation time $\tau^{-1}=\pi^{-1}n_{i}v_{0}^{2}\left(a^{2}+b^{2}\right)\int d^{2}\mathbf{q}\delta\left[\epsilon_{F}-\epsilon_{s}\left(\mathbf{q}\right)\right]$.
Although the Dirac delta function is explicitly spin dependent, the
integral over momentum is actually not. This is due to the special
band shape: $\epsilon_{+}\left(q,\theta\right)=\epsilon_{-}\left(q,-\theta\right)$.
$\tau$ is thus spin independent even if $\Sigma^{R,A}$ is solved
self-consistently, as will be shown in Supplemental Material Sec.
VII. The PHE percentage is defined by $\frac{\sigma_{\perp}\left(\theta_{{\cal E}}\right)}{\sigma_{\parallel}\left(\theta_{{\cal E}}\right)}\times100\%$,
which does not depend on $\tau$, $n_{i}$ or $v_{0}$. The numerical
result using the $k\cdot p$ model is shown in Fig. \ref{fig:Planner-Hall-effect}(c)
for $\epsilon_{F}=\epsilon_{V}-0.01\thinspace\textrm{eV}$. A major
result is the two-fold rotational symmetry of the PHE percentage,
rather than three-fold as suggested by the lattice. This $C_{2}$
symmetry originates from the $D_{2h}$ point group brought down from
$D_{6h}$ due to the magnetic anisotropy as discussed before. Particularly,
both the PHE percentage and the Hall conductivity $\sigma_{\perp}(\theta_{\varepsilon})$
are vanishing at $\theta_{{\cal E}}=n\pi/2$. At these angles, the
mirror reflection about the plane containing $\hat{z}$ and electric
field is a symmetry operation, which rules out the Hall effect.

\begin{figure}
\begin{centering}
\includegraphics[width=1\columnwidth]{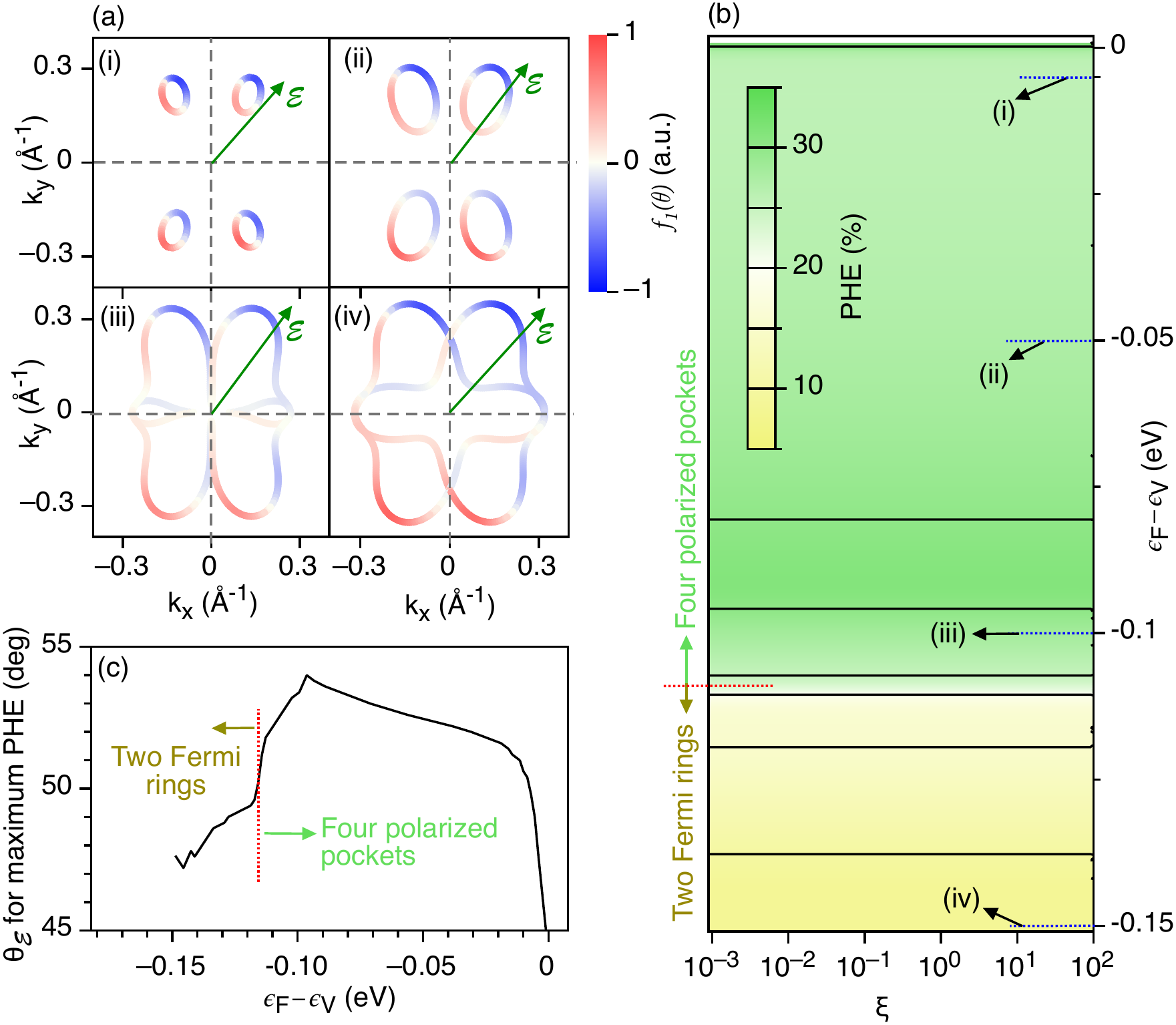} 
\par\end{centering}
\caption{Full-band Planar Hall effect (PHE) calculated from Boltzmann equation.
(a) The calculated nonequilibrim distribution function $f_{\mathbf{k}}^{1}$
along different Fermi circles. (b) The PHE percentage as a function
of $\epsilon_{F}$ and $\xi$. Here $\xi=\text{\ensuremath{\frac{N_{S}}{N_{C}}}}$is
the concentration ratio between spin-polarized and spin-unpolarized
scattering centers. (c) The current direction that can maximize PHE
percentage for different $\epsilon_{F}$ positions. \label{fig:Upper-limit-estimation}}
\end{figure}

The above transport analysis is based on the minimal $k\cdot p$ model
and a constant relaxation time. To show the quantitative accuracy,
this result is now compared to the one given by Boltzmann transport
equation (BTE) using the full \emph{ab initio} band. In
general, such an effort is necessary to handle the full anisotropic
band structure\citep{vyborny_semiclassical_2009,xiao_role_2017}.
Assuming uniform current distribution and a steady state, BTE is simplified
as $-e\boldsymbol{{\cal E}}\cdot\boldsymbol{\nabla}_{p}f=\frac{\partial f}{\partial t}\bigr|_{\textrm{coll}}$.
Here, $f\left(\mathbf{k}\right)=f_{0}^{\mathbf{k}}+f_{1}^{\mathbf{k}}$
is the total distribution function, and $f_{1}^{\mathbf{k}}$ denotes
the non-equilibrium part. Assuming $f_{1}^{\mathbf{k}}=g_{\mathbf{k}}\left(-\frac{\partial f_{0}}{\partial\epsilon_{F}}\right)\approx g_{\mathbf{k}}\delta\left(\epsilon-\epsilon_{F}\right)$,
detailed balance requires $-e\boldsymbol{{\cal E}}\cdot\mathbf{v}_{\mathbf{k}}=\sum_{\mathbf{k}'}\left(g_{\mathbf{k}}-g_{\mathbf{k}'}\right)S_{\mathbf{k}'\mathbf{k}}$,
where $S_{\mathbf{k}'\mathbf{k}}$ is the transition rate from $\mathbf{k}$
to $\mathbf{k}'$. Here, we consider two types of scattering mechanisms:
the spin-less Coulomb scattering and the exchange-induced spin-dependent
scattering. $\xi=\frac{N_{S}}{N_{C}}$, which is the concentration
ratio between these two types of impurities. Here the transition rate
sums over all considered scattering types, $S_{\mathbf{k}'\mathbf{k}}=\sum_{\alpha}S_{\mathbf{k}'\mathbf{k}}^{\alpha}$,
where $S_{\mathbf{k}'\mathbf{k}}^{\alpha}=\frac{2\pi N_{\alpha}}{\hbar}\bigl|H_{\mathbf{k}'\mathbf{k}}^{\alpha}\bigr|^{2}\delta\left(\epsilon_{k}-\epsilon_{k'}\right)$,
and $H_{\mathbf{k}'\mathbf{k}}^{\alpha}$ is the Hamiltonian that
scatters $\mathbf{k}$ to $\mathbf{k}'$. The scattering rate is evaluated
between the full-band eigenstates generated by Wannier90. With a bit
algebra, we obtain $g_{\mathbf{k}}=\tau_{0}^{\mathbf{k}}\bigl(\sum_{\mathbf{k}'}S_{\mathbf{k}'\mathbf{k}}g_{\mathbf{k}'}-e\mathbf{{\cal E}}\cdot\mathbf{v}_{\mathbf{k}'}\bigr)$,
where $\tau_{0}^{\mathbf{k}}=\bigl(\sum_{\mathbf{k}'}S_{\mathbf{k}'\mathbf{k}}\bigr)^{-1}$.
Note that the anisotropy of transport is fully absorbed by $g_{\mathbf{k}}$
without assuming a constant relaxation time. After discretizing the
Brillouin zone with a mesh of $250\times250$, $g_{\mathbf{k}}$ can
be solved through a linear system. The PHE percentage is defined as
$\textrm{PHE}=\frac{\sum_{k}f_{1}^{k}v_{\perp}^{k}}{\sum_{k}f_{1}^{k}v_{\parallel}^{k}}\bigr|_{\epsilon_{F}}\times100\%$,
which is compared to the $k\cdot p$ result given by Eq. \ref{eq:KuboFormula}
near the valence band edge {[}Fig. \ref{fig:Planner-Hall-effect}(c){]}.
Both transport models capture a $C_{2}$ rotation symmetry instead
of $C_{3}$. The solution of $f_{1}^{\mathbf{k}}$ at four different
$\epsilon_{F}$ positions (i-iv) are shown in Fig. \ref{fig:Upper-limit-estimation}(a).
A full scan of $\epsilon_{F}$ and the impurity concentration ratio
$\xi$ is then carried out, with the result shown in Fig. \ref{fig:Upper-limit-estimation}(b).
The PHE percentage is numerically estimated to be $25\%\sim31\%$
above the band crossing. Changing $\xi$ for several orders of magnitude
does not change this percentage, suggesting that the ratio between
the spin-dependent and independent scattering is not important. This
originates from the special band shape $\epsilon_{+}\left(q,\theta\right)=\epsilon_{-}\left(q,-\theta\right)$
as discussed before. Further details of this discussion can be found
in Supplemental Material Sec. VIII. The current
direction that maximizes PHE {[}$\theta_{{\cal E}}^{\textrm{max}}=\arg\max_{\theta_{{\cal E}}}\textrm{PHE}\left(\epsilon_{F}\right)${]}
is shown in Fig. \ref{fig:Upper-limit-estimation}(c) for different
$\epsilon_{F}$ positions. The value of $\theta_{{\cal E}}^{\textrm{max}}$
varies between $45^{\circ}\sim54^{\circ}$ for a wide range of energy,
which is determined by the details of the band shape. 

The $k\cdot p$ Hamiltonian given by Eq. \ref{eq:TheKP_Hamiltonian}
is not only effective, but also minimal. The quartic spin-orbit coupling
term is necessary, in the absence of which, the extra $C_{4}{\cal T}$
symmetry of the quadratic spin-orbit coupling term $\bigl(k_{x}k_{y}\sigma_{z}\bigr)$
rules out the Hall conductivity. \textcolor{black}{The $k_z=0$ approximation assumes 
that the transport is dominated by the first sub-band in a thin film of only a couple of 
unit cells. It is required to use a full bulk Hamiltonian to capture the crossover from 2D to 3D, which calls for future investigations.} 
The linear-system solution of BTE
used above is generic to include arbitrary combinations of elastic
scattering mechanisms. This allows for a single-step calculation of
the full-band non-equilibrium distribution without requiring self-consistent
iterations. The PHE percentage estimated by this calculation is more
than one order of magnitude greater than that observed in experiments,
suggesting a vast space to improve the device performance by engineering
the $\epsilon_{F}$, applying currents along $\theta_{{\cal E}}^{\textrm{max}}$,
or by scaling down the device to the limit of a single antiferromagnetic
domain. 

\textbf{Acknowledgements:} The transport analysis and simulations
done by G.Y., Y.L., R.L. and K.W. are supported by SHINES, an EFRC
funded by the US Department of Energy (DOE), Office of Science, Basic
Energy Sciences (BES) under award \#SC0012670. J.X.Y. and J.Z. were supported by the DOE of US, Office of Science, BES under Award No. DE-SC0016424. First-principles calculations were conducted on Extreme Science and Engineering Discovery Environment (XSEDE) under Grant No. TG-PHY170023, and Trillian, a Cray XE6m-200 supercomputer at UNH supported by the NSF MRI program under Grant No. PHY-1229408. G.Y. and K.W. are also grateful for
the support from National Science Foundation (DMR-1411085), and the
ARO program under contract W911NF-15-1-10561.

\bibliographystyle{apsrevNoURLNoISBN}

\end{document}